\documentclass[1p]{elsarticle}

\usepackage{lineno,hyperref}
\usepackage{graphicx, xcolor}
\usepackage{booktabs}
\usepackage{subcaption}
\usepackage{caption}
\usepackage{float}
\usepackage{import}
\usepackage{microtype}
\usepackage{amsmath}
\usepackage{verbatim}\usepackage{tikz}
\newcommand*\circled[1]{\tikz[baseline=(char.base)]{\node[shape=circle,draw,inner sep=0.1pt] (char) {#1};}}
\modulolinenumbers[5]

\journal{Journal of \LaTeX\ Templates}

\begin{document}
	
	\begin{frontmatter}
		
		\title{Analysis of process-induced damage in remote laser cut carbon fibre reinforced polymers}
		
		
		\author[IFKM]{Benjamin Schmidt}
		\author[IMS]{Michael Rose}
		\author[IMS]{Martina Zimmermann}
		\author[IFKM,DCMS]{Markus K\"astner\corref{mycorrespondingauthor}}
		\cortext[mycorrespondingauthor]{Corresponding author}
		\ead{markus.kaestner@tu-dresden.de}

		\address[IFKM]{Institute of Solid Mechanics, TU Dresden, Dresden, Germany}
		\address[IMS]{Institute of Materials Science, TU Dresden, Dresden, Germany}
		\address[DCMS]{Dresden Center for Computational Materials Science (DCMS), TU Dresden, Dresden, Germany}
		
		\begin{abstract}
			In this contribution a method is introduced that allows for a linkage between the  process-induced structural damage and the fracture behaviour.
			Based on an anisotropic elastic material model, different modelling approaches for initial damage effects are introduced and compared.
			The approaches are	applied to remote laser cut carbon fibre reinforced polymers in order to model various thermally induced damage effects like chemical decomposition, micro-cracks and delamination. The dimensions of this heat affected zone are calculated with 1D-heat conduction.
			In experiment and simulation milled and laser cut specimens with different process parameters are compared in order to quantify the impact of the cutting technology on the fracture behaviour. For this purpose open hole specimens were used.
		\end{abstract}
		
		
	\end{frontmatter}
	
	
	\section*{Nomenclature}
	\begin{table}[H]
		\small
		\centering
		\begin{tabular}{l p{7 cm}} 
			\toprule
			Latin characters & Description \\
			\midrule
			$d$ & damage variable\\
			$E$ & Young's modulus\\
			$f$ & material degradation factor \\
			$G$ & shear modulus\\
			$s$ & standard deviation \\
			$T$ & temperature\\
			$X$ & strength\\
			\midrule
			Greek characters & Description \\
			\midrule
			$\Gamma$ & fracture toughness\\
			$\varepsilon$ & strain \\
			$\mu$ & Mohr-Coulomb friction coefficient\\
			$\nu$ & Poisson's ratio \\
			$\lambda$ & heat conductivity \\
			$\rho$ & density \\
			$\sigma$ & stress \\
			$\tau$ & shear stress \\
			\midrule
			Subscripts & Description \\
			\midrule
			$\parallel$ & fibre direction\\
			$\bot$ & perpendicular to fibre direction \\
			$\bot\parallel $ & longitudinal direction\\
			$\bot\bot $ & traversal direction \\
			
			$C$ & compression \\
			$\mathrm{eff}$ & effective strain\\
			$\mathrm{fib}$ & fibre \\
			$\mathrm{init}$ & initial \\
			$\mathrm{kink}$ & fibre kinking\\
			$l$ & longitudinal direction in fracture plane \\
			$\mathrm{mat}$ & matrix \\
			$\mathrm{MEZ}$ & matrix evaporation zone \\
			$\mathrm{HAZ}$ & heat affectedzone \\
			$n$ & normal direction in fracture plane \\
			$T$ & tension \\
			$t$ & traversal  direction in fracture plane \\
			\bottomrule
		\end{tabular}
	\end{table}
	
	\section{Introduction}
	During the manufacturing process, especially cutting processes of structures, damage, e.g. micro-cracks, internal stresses or material degradation, may occur. Often those damage effects are unidentified and thus not regarded in the structural analysis, even if they can have a relevant impact on the fracture behaviour.\\
	The aim of this paper is the analysis of process-structure-property linkages between 
	process-induced damage in anisotropic materials and their influence on the fracture behaviour. The introduced modelling approach is applied to thermal damage in remote laser cut Carbon Fibre Reinforced Polymers (CFRP). \\
	A common technology for CFRP is mechanical processing, during which high tool wear occurs due to the high strength of the carbon fibres. The cutting quality decreases with the increasing wear of the cutting tool. Laser beam cutting, on the other hand, works without force or contact, so that no mechanical tool wear occurs. The cutting quality is constant. When processing CFRP, remote laser cutting has advantages over gas-assisted laser cutting. The higher spot speeds allow for a short interaction time between the material and the laser beam. This reduces the heat input into the surrounding material. \\ 
	Caused by the heat input, pores and matrix degradation occur in the area close to the cutting gap. A review of this Heat Affected Zone (HAZ) and  its damage phenomena can be found in \cite{herzog08}.
	In the present paper, the temperature field during the laser cutting process is modelled with 1D thermal heat conduction. Based on this temperature field the extent of the HAZ is calculated for different cutting configurations. Thermally induced material degradation in this area is modelled as a reduction of mechanical material parameters, which allows for a linkage between the cutting process and the mechanical properties of the cut material.
	As an example, comparative tensile tests of milled and laser cut open hole specimens are simulated. The advantage of this specimen type is the predefined localisation of damage at the cutting edge of the holes, which allows a direct comparison of the cutting technologies due to the controlled heat input at the failure initiation zone around the hole.\\
	For the modelling of the described thermally induced damages in CFRP, two different approaches were developed and compared. In the first, the damage variables of the material model are utilised, the second approach uses a material parameter reduction.
	
	\section{Remote laser cut CFRP}
	\subsection{Process description}
	In remote laser beam cutting, the laser beam is deflected into the processing area by two galvanometrically driven tilting mirrors. Each mirror controls the position of the laser spot in one Cartesian axis, which makes it possible to represent any contour to be cut. Depending on the system configuration, the light, dynamic mirrors allow spot speeds of up to several meters per second on the material surface. Unlike gas-assisted laser beam cutting, no gas flow is required to remove the waste products from the cutting gap volume. The gas pressure generated during the thermal sublimation drives particulate, molten and gaseous waste products out of the cutting gap.
	The high spot speeds of remote laser beam cutting result in a shorter interaction time between the material to be cut and the laser beam compared to gas-assisted laser beam cutting. In this way, it is possible to reduce the HAZ at the cutting edge. Furthermore, when cutting fibre composites in general and CFRP in particular, several exposure cycles of the cutting contour are required until the cutting gap is completely formed by the laser. Pause times before the repeated intervention of the laser beam lead to a cooling of the process zone and further improve the cutting quality \cite{klotzbach2011laser}.
	\begin{figure}[!h]
		\def\svgwidth{\columnwidth}	
		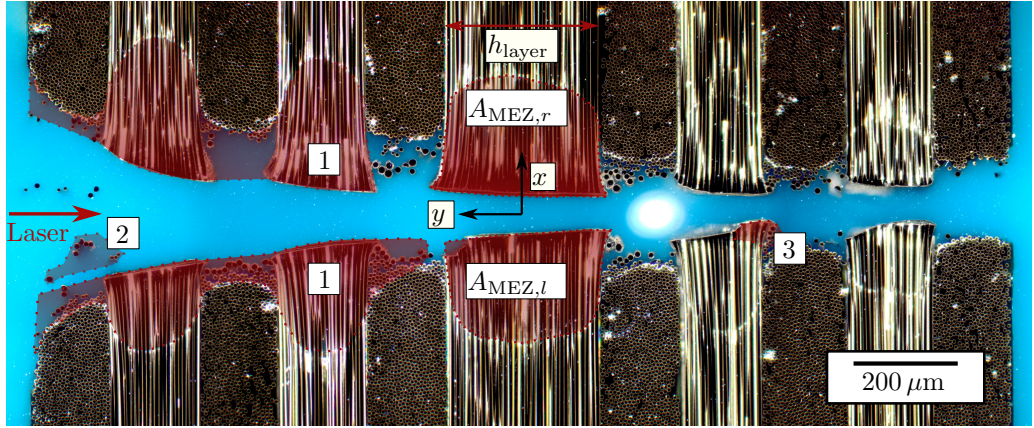
		\caption{Micro section of remote laser cutting gap with visible damage effects: (1) matrix evaporation zone (MEZ), (2) detached parallel fibres, (3) thermally damaged fibres (volume increase)}
		\label{fig:Schliff}
	\end{figure}
	
	\subsection{Heat affected zone}
	Laser cutting is a thermal cutting process. Due to its thermal conductivity, the material around the cutting edge is heated during the process. As a result, the HAZ is formed. The matrix evaporation zone (MEZ) is presented as a part of the HAZ in \autoref{fig:Schliff}. The inhomogeneous material structure of CFRP is reflected in different orders of magnitude of the decomposition temperatures of the composite components. Carbon fibres sublimate at about $3900 \, \mathrm{K}$ \cite{weber2011minimum}, the 
	sublimation temperature of a polymer such as epoxy is about $800 \, \mathrm{K}$ \cite{klotzbach2011laser}. This circumstance causes areas 
	at the cutting edge where the polymer matrix has already evaporated because its sublimation temperature was exceeded, while carbon fibres are still present. As the distance to the cutting edge increases, the maximum process temperatures decrease. In areas where the matrix evaporation temperature was not exceeded, still pyrolysis may occur. This thermal decay of the polymer matrix system is also part of the HAZ.\\
	In \cite{young2008} the mechanical damage caused during micro hole laser drilling ($d=120 \, \mathrm{\mu m}$) in CFRP is analysed. Here, drilled specimens are compared to intact specimens in static and cyclic tensile tests. As a result, a $10 \, \%$ stiffness reduction and $29 \, \%$ strength reduction were measured during the static tests. An application of these values for material modelling is not recommended, since the influence of the heat affected zone is hard to separate from the additional stress concentration at the holes.\\
	Stock et al.~\cite{stock2016} examined the stress concentration in open hole tensile tests of woven CFRP. For the static case, the stress concentration around the hole is calculated analytically. In the comparison of laser cut and water jet cut specimen a stiffness reduction in the HAZ was observed, which lead to decreased stress peaks at the holes and thus to higher maximum forces for laser cut specimens.\\
	Open hole specimens were also used in \cite{kalyanasundaram2018} in order to examine the HAZ. Here cyclic tensile tests on different stress levels were performed. Kalyanasundaram et al. observed increasing damage with increasing stresses while at lower stress levels a sudden failure occurred. Similarly at higher stress levels a higher stiffness reduction was determined before fracture.\\
	In \cite{weber2011minimum} the location of different damage phenomena that occur during the laser cutting process is analysed by modelling heat conduction and identifying the maximum distance from the cutting edge where different characteristic temperatures occur. These temperatures e.g. the matrix evaporation temperature allow for the identification of the dimensions of the HAZ.\\
	Hejjaji et al. \cite{hejjaji2016machining} compared conventional drilling and fibre laser machining in carbon and glass fibre reinforced plastics. The damage that occurs for both cutting technologies was measured using scanning electron microscopy and scanning acoustic microscopy. In comparison of the technologies, the dominant damage in conventional cutting is exit-ply lamination, whereas in fibre laser machining the HAZ is identified. In conventional drilling a better cutting surface quality was reached.
	Tao et al. \cite{tao2020dual} presented a laser cutting technology for thick ($d = 10 \, \mathrm{mm}$) CFRP plates based on two laser beams, one from the up- and one from the downside. The resulting HAZ was analysed and the process modelled with a thermal FEM model. It could be shown that the dual-beam approach could minimise the HAZ. The analysis of the cutting process showed, that depending on the hole depth the process can be divided in three stages, with increasing depth the cutting process slows down.\\
	Xu et al. \cite{xu2019effects} examined thermal ageing in carbon fibre reinforced epoxy composites with ageing cycles of up to $190 \, ^\circ\mathrm{C}$. Open hole tests were performed to compare intact and aged specimens. Here for all temperatures and ageing times the tensile strength was increased compared to the initial specimen and a maximum strength was found for $t=250 \, \mathrm{s}$\\
	Chippendale et al. \cite{chippendale2014numerical} describe a modelling approach for the thermal decomposition due to the laser heat input with a focus on the influence on the thermal conductivity. The main degradation effects were identified as polymer pyrolysis to chars and gases and carbon fibre sublimation. The damage in the laminates observed by Chippendale were much worse and on a larger scale than those caused by remote laser cutting, since in remote laser cutting there are used much higher laser powers and spot velocities, but the exposure time is only a fraction in remote laser cutting.\\
	While Chippendale assumes the degradation to be dependent on the interaction time and thus uses an Arrhenius approach for the degradation model for the polymer, another approach is introduced in Canisius 2018 \cite{canisius2018}: there the matrix is assumed to sublimate as well as the fibre after the sublimation temperature is reached and the decomposition energy is absorbed.

	\subsection{Damage and failure of CFRP}
	The analysis of the influence of the HAZ on the mechanical properties requires a precise model for the behaviour of the initially intact material. Based on this model the thermal damage can be modelled. For this purpose, different failure models and a suitable specimen geometry are presented.
	\paragraph{Failure criteria}
	In modelling the failure behaviour of CFRP, there are two main challenges. On the one hand side, caused by the heterogeneous material structure, there is a manifold of different failure modes, that need to be regarded. On the other hand side, the scale on which failure happens is very small compared to the size of the structure. A possible solution for the latter challenge is the use of macroscopic material models. Here, fibre and matrix are layer-wise replaced by an effective homogeneous transversely isotropic material. Similarly, to account for damage and failure not the micro-cracks, but the reduction of the macroscopic material properties is modelled.
	In general, failure criteria for CFRP are micro-mechanically motivated, divided in fibre and matrix dominated failure modes.\\
	Hashin \cite{hashin1980failure} developed three-dimensional failure criteria in terms of quadratic stress polynomials and presented a first criterion for matrix tensile failure with an action plane.
	Puck \cite{puck2002failure} first applied a Mohr-Coulomb approach for matrix compression failure, which was later improved by Cuntze et al. \cite{cuntze2004predictive} with a combination of nine individual failure modes.
	\\
	For fibre modes different failure criteria were introduced.
	Fibre tensile failure is widely assumed to be independent from the other modes, therefore the failure criterion depends only on the fibre tensile strength. For fibre compression there are interactions to the other modes. D\'avila et al. \cite{davila2003failure}  described kinking of the fibres under pressure and presented a two-dimensional fibre kinking model, which was later advanced to a three-dimensional formulation by Pinho et al. \cite{Pinho06_PartI} and \cite{Pinho06_PartII}. In \cite{maimi2007continuum} a similar approach is presented with different numerical adaptations e. g. mesh size independency for energy release rates.\\
	If the partial degradation of the material in the HAZ is of interest, e.g. to evaluate the resulting specimen strength, the concept of continuum damage mechanics \cite{lemaitre1996course} can be used to represent the individual laminae. For isotropic damage phenomena, simple models with a single scalar variable are common approaches. Anisotropic formulations that distinguish characteristic failure modes, e.g. matrix and interface degradation as well as fibre failure, are available in the literature. For the two-dimensional case Matzenmiller et al. \cite{matzenmiller1995constitutive}	presented an anisotropic damage model where damage initiation is controlled by the failure criteria developed by Hashin \cite{hashin1980failure}. Chatiri and Matzenmiller \cite{chatiri2013damage} extend this model to the three-dimensional case and incorporate damage initiation based on the Puck failure criterion.\\
	An extensive overview of different inter-fibre damage effects and modelling approach for the shear plasticity can be found in \cite{mandel2017mechanism}. Based on Puck's failure criterion, Schirmaier et al. \cite{schirmaier2014new} developed a new approaches for the determination of the fracture angle, which accelerated the finite element analysis significantly. Wei et al. \cite{wei2019new} introduced a continuum damage model with an improved in-plane shear damage model that distinguished two different shear modes.\\
	An anisotropic damage  model with separate damage variables for fibre tensile, fibre compression and matrix mode is presented by Pinho et al. in \cite{Pinho06_PartI} and \cite{Pinho06_PartII}.
	\\
	In order to compare different models and to evaluate their predictions, the world wide failure exercise was organised. Here the failure model's predictions were compared to experimental results for different modes. In \cite{kaddour2013maturity} the results of this test are presented. The damage model by Pinho et al. \cite{Pinho06_PartI} achieved excellent results, in the qualitative assessment as well as in the quantitative. It is therefore used in this contribution.
	\paragraph{Open Hole Test}
	In testing and development of composites, the open hole test specimen is widely used \cite{stock2016,chen2013numerical,lee2015initial,mohammadi2017correlation,yoon2019development,zhang2019damage,almeida2020improving}. It allows for the comparison of machining technologies and during failure a combination of different damage effects like delamination, matrix cracks and fibre breakage occurs.\\
	Chen et al. \cite{chen2013numerical} analysed the influence of the element size in the simulation of open hole tensile test specimens and presented a layer thickness depending approach for the fracture toughness.\\
	In \cite{lee2015initial}, based on Puck's failure criterion, a coupled failure initiation and progression was introduced and applied to open hole tensile test. The results achieved a good agreement to experimental results.
	Mohammadi et al. \cite{mohammadi2017correlation} examined open hole tensile test using Acoustic Emission and Scanning Electron Microscope to identify matrix cracking, matrix/fibre debonding and fibre breakage and validated the results with an FE analysis. In \cite{yoon2019development} a new modelling approach for fracture toughness normalisation is introduced, which reduced mesh dependency.
	Zhang et al. \cite{zhang2019damage} simulated open hole tension and compression tests regarding shear non-linearity and fibre kinking, which showed good agreement to experimental results.
	Almeida et al. \cite{almeida2020improving} presented a numerical and experimental study on the optimisation of open hole structures using tailored fibre placement.

	\section{Experimental and numerical analysis}
	The experimental and numerical analysis consists of two different parts. First the initially intact CFRP is characterised with tensile tests and a damage model for this material is presented, parametrised and validated. On this basis the HAZ is analysed in the second part. Different cutting parameter configurations are compared and a method for the calculation of the extent of the HAZ is presented. Remote laser cut tensile test specimens are experimentally tested and simulated with two different modelling approaches for the HAZ.
	\subsection{Experimental analysis}
\begin{figure}[!h]
		\def\svgwidth{\columnwidth}	
		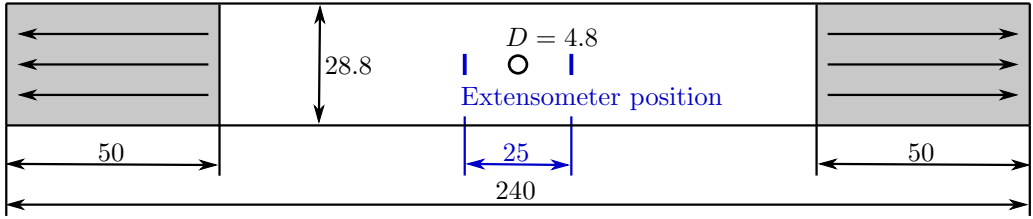
		\caption{Geometry of the open hole specimen, dimensions in $[\mathrm{mm}]$. Grey rectangles show the clamping area.}
		\label{fig:Probekoerper}
	\end{figure}

	\paragraph{Description of the material}
	The investigated composite material is a carbon fibre reinforced epoxy. The laminates were pressed from Sigrapreg C U150-0/NF-E340/$38\, \%$ prepregs. The press pressure for the material consolidation was $392 \, \mathrm{kPa}$. Within three hours, the press was heated up from room temperature to $110\, ^\circ \mathrm{C}$ and then cooled down slowly within 13 \, hours. The fibre type is Torayca T700 without an additional fixation of the rovings (NF) and the epoxy system is E340 with a glass transition temperature of $T_G=140 \, ^\circ\mathrm{C}$. The areal weight of the prepregs is $150 \, \mathrm{g m}^{-2}$ with a relative mass content of the epoxy resin of $38 \, \%$. For the mechanical characterisation of the material, two configurations were produced, which are described in more detail in \autoref{tab:MatConfig}. Fibre volume content and mass density were determined according the standards DIN EN 2564 (wet-chemically) and  DIN EN ISO 1183-1, respectively.
	\begin{table}[H]
		\centering
		\begin{tabular}{lrr} 
			\toprule
			Configuration & Uni-Directional (UD) & Multi-Layer (ML) \\
			\midrule 
			Stacking sequence & $[0]_{14}$ & $[0/90]_{6\mathrm{s}}$\\
			Thickness in $\mathrm{mm}$ & 2.2 & 1.9\\
			Measured fibre volume content & $54\, \%$& $54\, \%$\\
			\bottomrule
		\end{tabular}
		\caption{Material Specifications}
		\label{tab:MatConfig}
	\end{table}
	\paragraph{Open hole test specimen}
	In order to identify the influence of the laser cutting process on the fracture properties of CFRP, a suitable specimen type is needed. In unnotched laser cut specimens the HAZ is located at the outer contour, but failure initiates at random spots all over the specimen because of material imperfections, therefore this type is not appropriate. \\
	A better suited specimen type is the open hole specimen. Through the laser cut holes, the HAZ is located in the highly stressed area around the hole. 
	With this overlap of stress concentration and HAZ, a high sensitivity of the failure behaviour to the thermal degradation is reached.\\
	The geometry of the open hole test specimen is given in \autoref{fig:Probekoerper}.
	The outer contour of the open hole specimens is milled, a hole inside the specimen is either milled, so assumed initially intact and used as a reference, or laser cut with different cutting configurations given in \autoref{tab:CutConf1}. For the milled contours a rotational speed of $24 \, 000 \, \mathrm{min}^{-1}$ and a feed rate of $900 \, \mathrm{mm} \; \mathrm{min}^{-1}$ were used. In a micro section of the milled cutting edge (\autoref{fig:VglCutConf}) no damages like fibre pull-out, matrix-cracks and delaminations were visible.\\
	Displacements were measured with a tactile extensometer which was placed in the middle of the specimen over the hole with a reference length of $25 \, \mathrm{mm}$.

	\paragraph{Cutting process parameters}
	In order to analyse the influence of the HAZ on the mechanical properties, three different cutting parameter configurations with different extents of the HAZ are tested. The first two configurations use a fibre laser with different power and spot velocity, the third uses a CO$_2$ laser. The full cutting parameters are given in  \autoref{tab:CutConf1}. For each configuration a micro section  is produced, which is used for the determination of the MEZ, see \autoref{fig:VglCutConf}. Here it showed, that variations of the cutting parameters significantly impacted the size of the gap and the MEZ. The impact of the cutting parameters on the mechanical behaviour is analysed in the following sections.
	\begin{table}[!h]
		\centering
		\begin{tabular}{p{3.5cm} rrrr} 
			\toprule
			Parameter & Unit & L12 & L25 & CO$_2$ \\
			\midrule
			Beam source                    & $-$          &fibre laser&fibre laser& CO$_2$ laser\\
			Wavelength                     & $\mu \mathrm{m}$      &1.07&1.07 &10.6\\
			Laser Power                    & $\mathrm{W}$          &1860& 4650& 2870\\
			Focal spot diameter            & $\mathrm{\mu m}$      &36  & 36  & 482\\
			Intensity distribution         & $-$                   & Gaussian& Gaussian& Gaussian\\
			Spot velocity                  & $\mathrm{m \, s}^{-1}$&  5 &  1  &  0.5\\
			Cool down time                 & $\mathrm{s}$          &  1 &  1  &   2\\
			cutting cycles                 & $-$                   & 40 &  4  &   7 \\
			\bottomrule
		\end{tabular}
		\caption{Parameters of three different Laser cutting Configuration.}
		\label{tab:CutConf1}
	\end{table}
		
	\subsection{Continuum damage model}
	\label{par:Pinho}
	
\begin{figure}[H]
		\centering
		\def\svgwidth{0.45 \columnwidth}	
		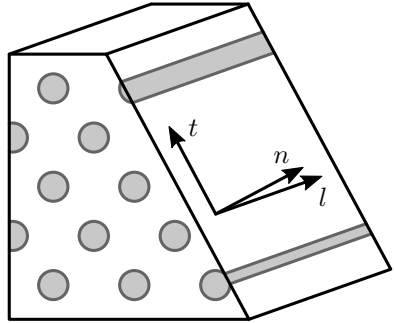
		\caption{Action plane for matrix failure and fibre kinking}
		\label{fig:ActPlane}
	\end{figure}

	To capture the characteristic damage phenomena in CFRP structures, the orthotropic continuum damage model introduced by Pinho et al. \cite{Pinho06_PartI} is utilised. The definition of the material model comprises three fundamental ingredients:\\
	\begin{enumerate}
		\item a stress--strain relation that accounts for damage in the material,
		\item failure criteria if an evolution of damage may occur, and
		\item evolution equations that define the increase of the damage variables.
	\end{enumerate}	
	The material model is based on physically motivated failure criteria for fibre tensile, fibre kinking (compression), matrix tensile and matrix compression modes, which are collected in \autoref{tab:FailCritPinho}.
	For brevity we only present the used constitutive relations without details of their derivation. A more	comprehensive presentation of the damage model can be found in \cite{Pinho06_PartI} and \cite{Pinho06_PartII}.	
	\begin{table}[H]
		\centering
		\begin{tabular}{p{2 cm}rl}
			\toprule
			Criterion & \multicolumn{2}{l}{Equation}\\
			\midrule
			Fibre tension & $f_{\parallel, \mathrm{T}}=$&$\frac{\sigma_\parallel}{X_{\parallel,\mathrm{T}}}-1$  \vspace{2 mm}\\
			Fibre kinking & $f_\mathrm{kink} =$&$ \begin{cases} \Big(\frac{\tau_t}{X_{\bot \bot}-\mu_t \sigma_n}\Big)^2 +\Big(\frac{\tau_l}{X_{\bot \parallel} - \mu_l \sigma_n}\Big)^2 -1 \Big| \sigma_n \le 0 \\
			\Big(\frac{\sigma_n}{X_\bot}\Big)^2+\Big(\frac{\tau_t}{X_{\bot \bot}}\Big)^2+\Big(\frac{\tau_l}{X_{\bot \parallel}}\Big)^2 -1\Big| \sigma_n > 0\end{cases}$ \vspace{3 mm}\\
			Matrix tension & $f_{\bot, \mathrm{T}} =$&$ \Big(\frac{\sigma_n}{X_\bot}\Big)^2+\Big(\frac{\tau_t}{X_{\bot \bot}}\Big)^2+\Big(\frac{\tau_l}{X_{\bot \parallel}}\Big)^2-1$\vspace{2 mm}\\
			Matrix compression & $f_{\bot, \mathrm{C}} =$&$\Big(\frac{\tau_t}{X_{\bot \bot}-\mu_t \sigma_n}\Big)^2 +\Big(\frac{\tau_l}{X_{\bot \parallel} - \mu_l \sigma_n}\Big)^2 -1$\\
			\bottomrule
		\end{tabular}
		\caption{Failure criteria in the damage model, failure initiates for $f=0$. $\{\sigma_n, \tau_t, \tau_l\}$ are the stress components in the action plane (see \autoref{fig:ActPlane}), $X$ is the strength and $\mu$ the Mohr-Coulomb friction coefficient. 
		}
		\captionsetup{width=0.4\linewidth}
		\label{tab:FailCritPinho}
	\end{table}
	Here, different failure criteria are noted according to the action plane presented in \autoref{fig:ActPlane} with the in-plane shear stresses $\tau_t$ (perpendicular to fibre) and $\tau_l$ (parallel to fibre) 
	.\\
	\begin{figure}[H]
		\centering
		\def\svgwidth{0.45\columnwidth}	
		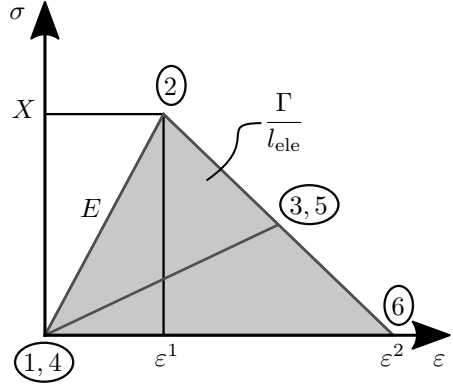
		\caption{Schematic representation of the uniaxial stress-strain curve in the presented damage model}
		\label{fig:Degrad}
	\end{figure}
	As presented in \autoref{fig:Degrad}, after failure initiation \circled{2} a linear damage evolution
	\begin{equation}
	d=\mathrm{max} \left\{ 0,\mathrm{min} \left\{1, \frac{\varepsilon-\varepsilon^1} {\varepsilon^2-\varepsilon^1} \right\}\right\}
	\end{equation}
	is assumed. For a material with a given damage state $\circled{3}$, the remaining stiffness, strength and fracture toughness are predefined, as presented in \autoref{fig:Degrad}: the initial stiffness $E_{\circled{1} \circled{2}}$ is reduced to $E_{\circled{4}\circled{5}}$, the remaining strength is reduced from $X_{\circled{2}}$ to $X_{\circled{3}}$ and the fracture toughness is reduced from the triangle area $\circled{1}-\circled{2}-\circled{6}$ to $\circled{4}-\circled{5}-\circled{6}$. \\
	The stresses are degraded with separate damage parameters for fibre tensile $d_\parallel$, fibre compression $d_\mathrm{kink}$ and matrix mode $d_\bot$ as presented in \autoref{tab:DamageVariables}. While for fibre compression and matrix mode only the stress components in the action plane are reduced, a catastrophic failure is assumed for the fibre tensile mode. Therefore, all stress components are degraded.
	The slope of the degradation curve is defined by the fracture toughness $\Gamma$; the combined area under the elastic stress-strain curve and the degradation curve is equal to the normalised fracture toughness. This normalisation with the element length $l_\mathrm{ele}$ perpendicular to the fracture plane was introduced to ensure mesh independence.\\
	\begin{table}[!h]
		\centering
		\begin{tabular}{rl}
			\toprule
			Mode & reduction formulation \\
			\midrule
			Fibre tensile & $ \sigma_{i} \leftarrow \Big(1-d_\parallel \frac{\left< \sigma_{i,\mathrm{init}}\right>}{\sigma_{i,\mathrm{init}}} \Big) \sigma_{i,\mathrm{init}} \; | \; i\in \{a,b,c\}  $ \\
			& $ \{\tau_{ab},\tau_{bc}, \tau_{ca}\} \leftarrow (1-d_\parallel ) \{\tau_{ab},\tau_{bc}, \tau_{ca}\}_\mathrm{init}  $ \\
			\midrule
			Fibre kinking & $ \sigma_n \leftarrow \Big(1-d_\mathrm{kink} \frac{\left< \sigma_{n,\mathrm{init}}\right>}{\sigma_{n,\mathrm{init}}} \Big) \sigma_{n,\mathrm{init}}  $ \\
			& $ \tau_t \leftarrow (1-d_\mathrm{kink} ) \tau_{t,\mathrm{init}}  $\\
			& $ \tau_l \leftarrow (1-d_\mathrm{kink} ) \tau_{l,\mathrm{init}}  $\\
			\midrule
			Matrix mode & $ \sigma_n \leftarrow \Big(1-d_\bot \frac{\left< \sigma_{n,\mathrm{init}}\right>}{\sigma_{n,\mathrm{init}}} \Big) \sigma_{n,\mathrm{init}}  $ \\
			& $ \tau_t \leftarrow (1-d_\bot ) \tau_{t,\mathrm{init}}  $\\
			& $ \tau_l \leftarrow (1-d_\bot ) \tau_{l,\mathrm{init}}  $\\
			\bottomrule
		\end{tabular}
		\caption{Reduction of the initial stresses with damage variables for the three modes}
		\captionsetup{width=0.4\linewidth}
		\label{tab:DamageVariables}
	\end{table}
	For the modelling of macroscopic cracks, element deletion is utilised. Elements are either deleted after reaching a damage value of $d=1$ for one of the damage modes or with the strain criterion reaching the effective failure strain $\varepsilon_\mathrm{eff}=\sqrt{\varepsilon_{ij} \varepsilon_{ij}}$.
	All in all, the material parameters that need to be identified in experiments are the orthotropic stiffnesses and Poisson's ratios and for the different failure modes the strengths and fracture toughnesses. The described material model is implemented in LS-Dyna \cite{LS-Dyna2020}.

\begin{figure}[!h]
	\centering
	\begin{subfigure}[b]{0.49\textwidth}
		\def\svgwidth{\columnwidth}	
		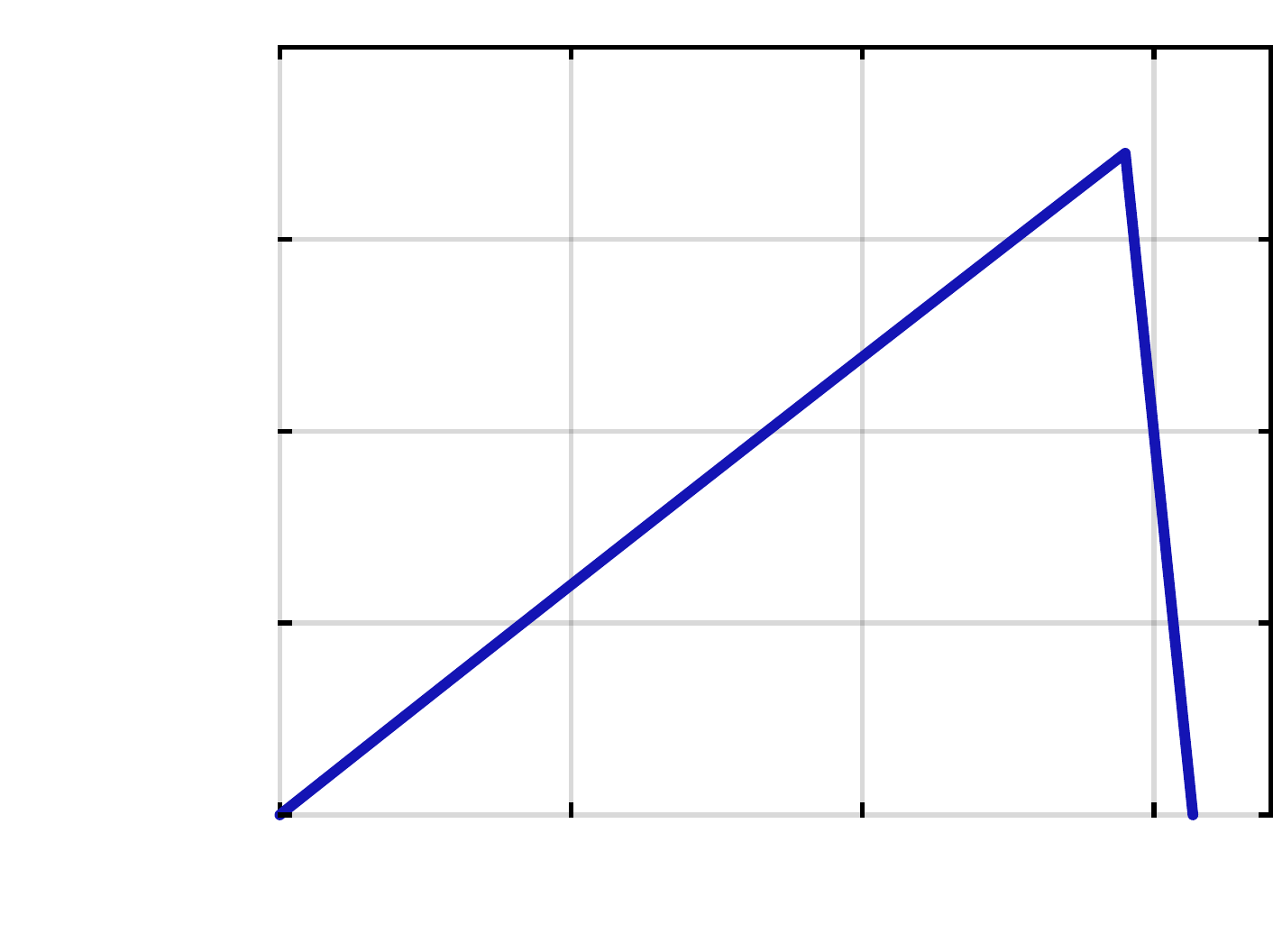
		\caption{Uni-directional $0^\circ$}
		\label{fig:UD0}
	\end{subfigure}
	\begin{subfigure}[b]{0.49\textwidth}
		\def\svgwidth{\columnwidth}	
		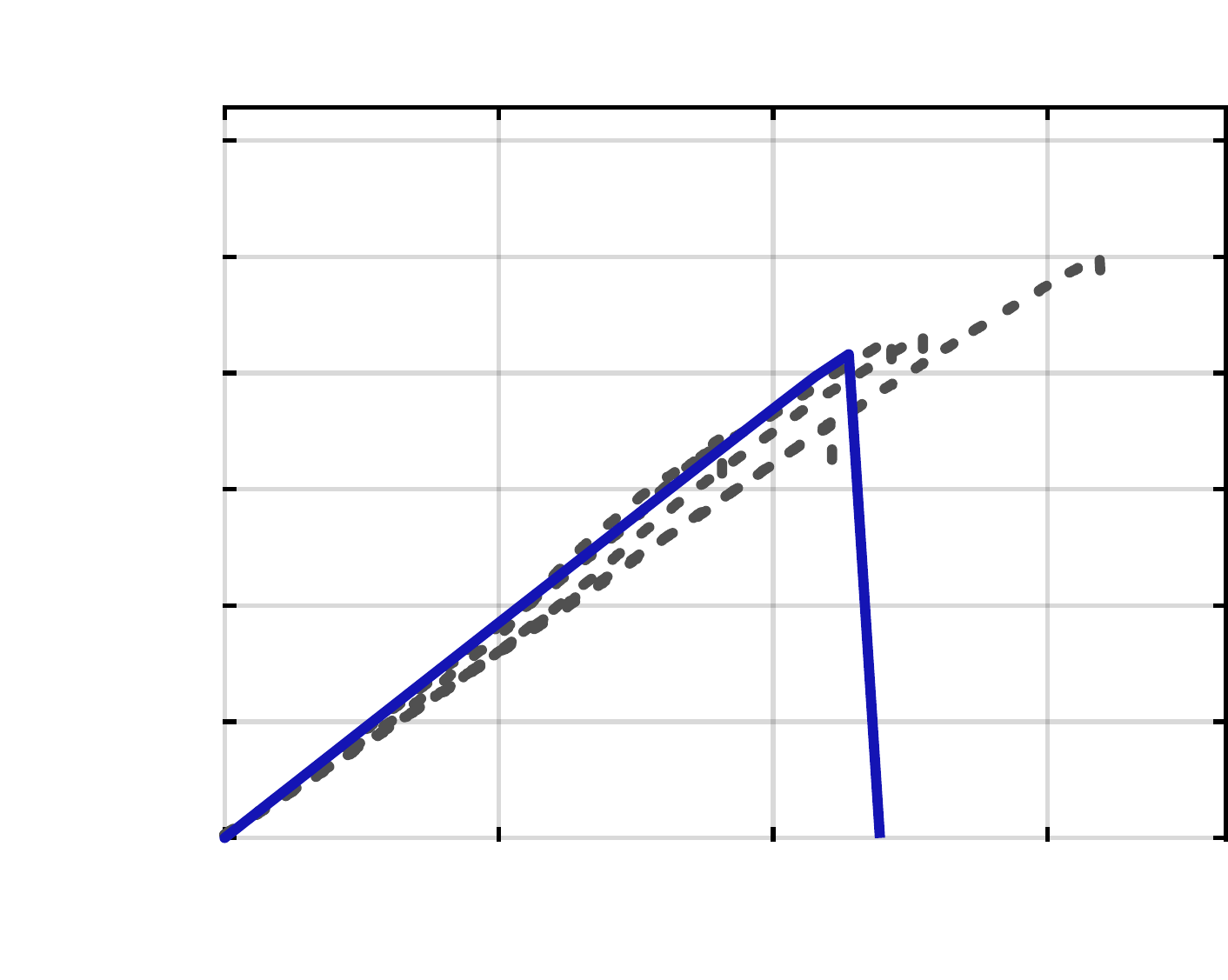
		\caption{Uni-directional $90^\circ$}
		\label{fig:UD90}
	\end{subfigure}
	\begin{subfigure}[b]{0.49\textwidth}
		\def\svgwidth{\columnwidth}	
		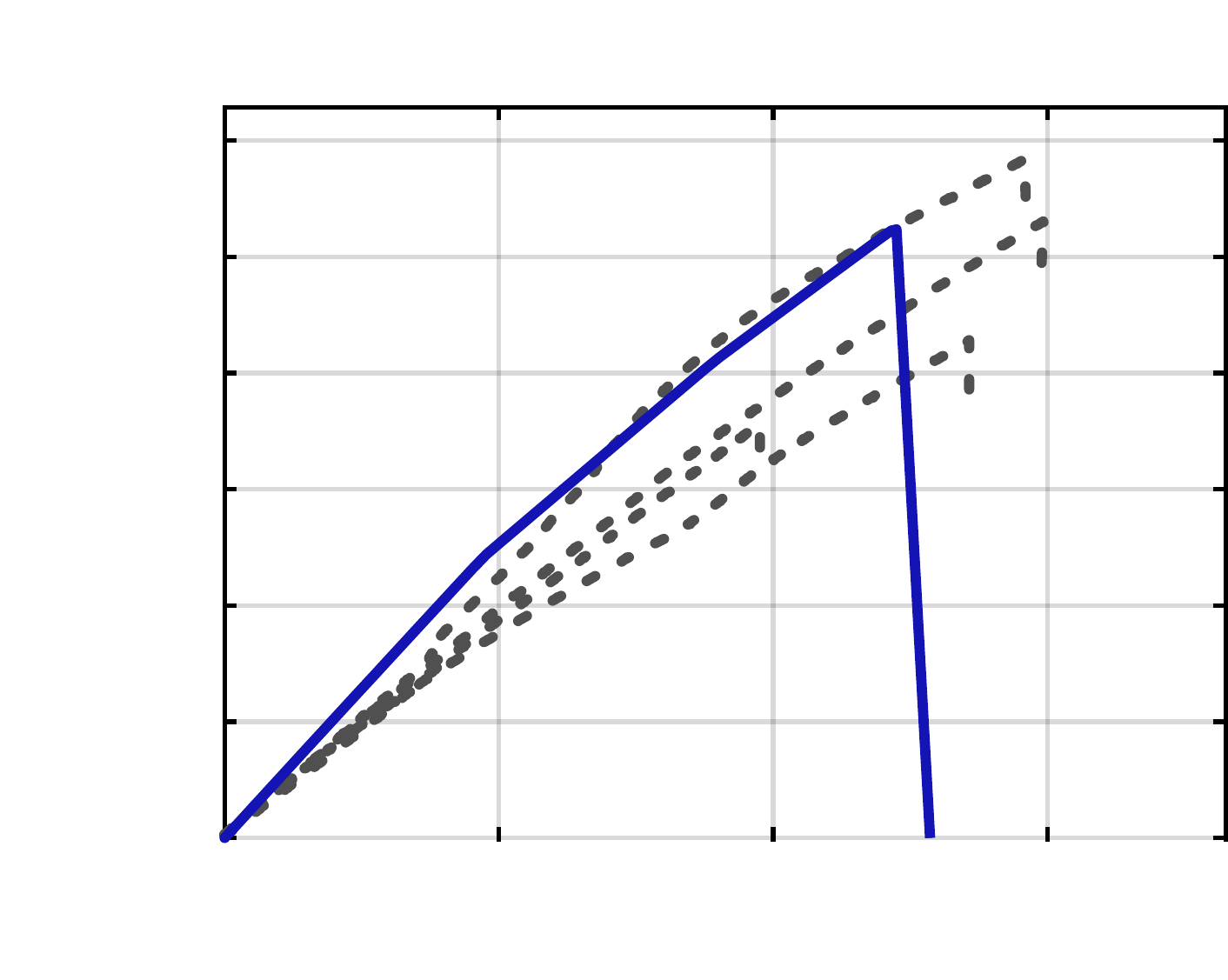
		\caption{Uni-directional $45^\circ$}
		\label{fig:UD45}
	\end{subfigure}
	\begin{subfigure}[b]{0.49\textwidth}
		\def\svgwidth{\columnwidth}	
		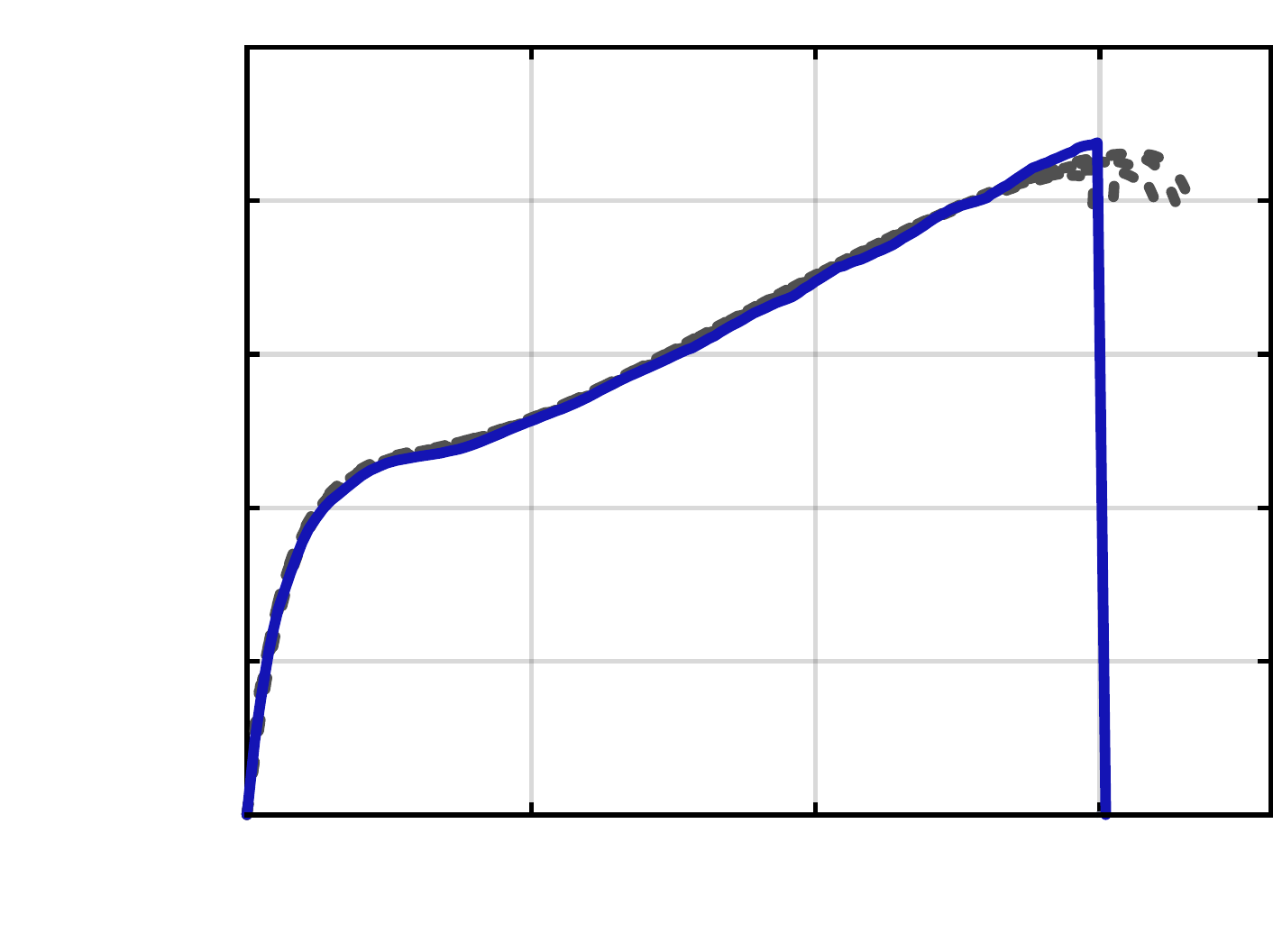
		\caption{Multi layer $\pm 45^\circ$}
		\label{fig:ML+-45}
	\end{subfigure}
	\begin{subfigure}[b]{0.49\textwidth}
		\def\svgwidth{\columnwidth}	
		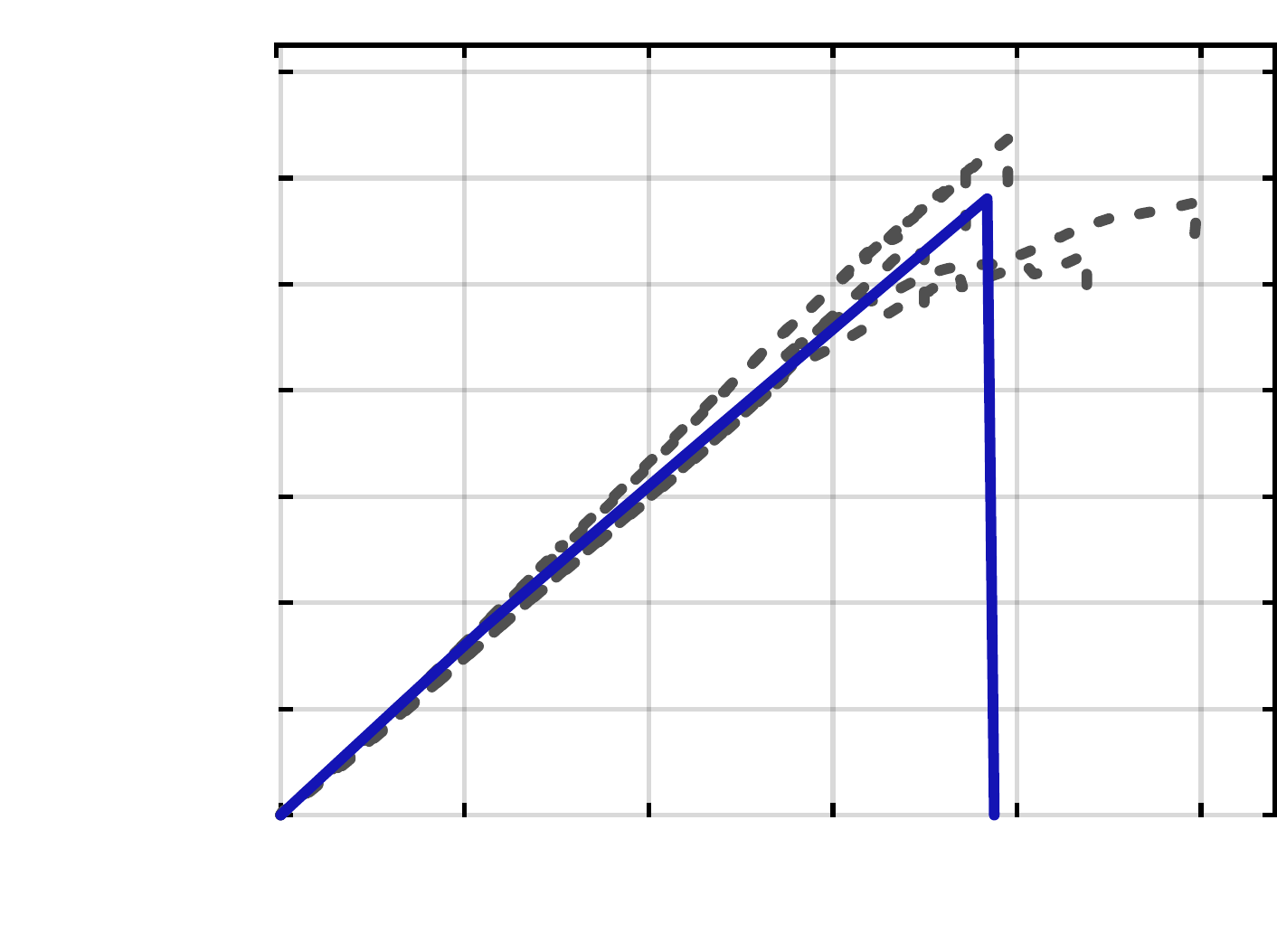
		\caption{Prediction: multi layer $0^\circ$/$90^\circ$}
		\label{fig:ML0-90}
	\end{subfigure}
	\caption{Stress-strain-relation for different material directions}\label{fig:KW_all}
	\label{fig:Paramet}
\end{figure}

\begin{figure}[!h]
	\centering
	\begin{subfigure}[b]{\textwidth}
		\includegraphics[width=\textwidth]{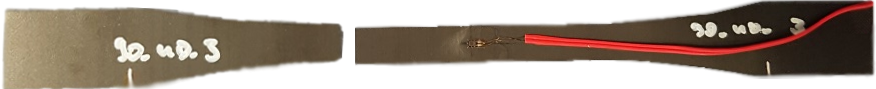}
		\caption{Experiment $90 \, ^\circ$-UD}
		\label{fig:BB_90_exp}
	\end{subfigure}
	\begin{subfigure}[b]{\textwidth}
		\includegraphics[width=\textwidth]{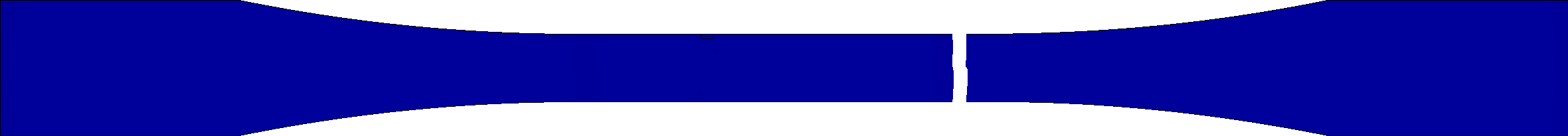}
		\caption{Simulation $90 \, ^\circ$ UD}
		\label{fig:BB_90_sim}
	\end{subfigure}
	\begin{subfigure}[b]{\textwidth}
		\includegraphics[width=\textwidth]{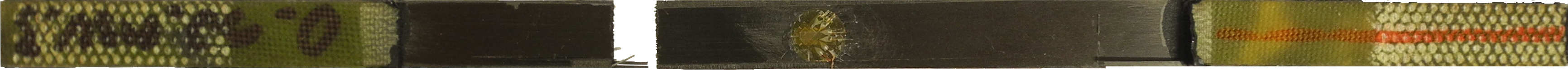}
		\caption{Experiment $0 \, ^\circ/90 \, ^\circ$ ML}
		\label{fig:BB_0-90_exp}
	\end{subfigure}
	\begin{subfigure}[b]{\textwidth}
		\includegraphics[width=\textwidth]{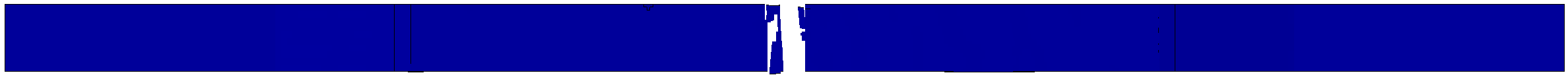}
		\caption{Simulation $0 \, ^\circ/90 \, ^\circ$ ML}
		\label{fig:BB_0-90_sim}
	\end{subfigure}
	\begin{subfigure}[b]{\textwidth}
		\includegraphics[width=\textwidth]{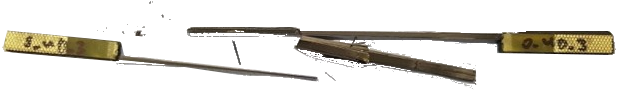}
		\caption{Experiment $0 \, ^\circ$ UD}
		\label{fig:BB_0_exp}
	\end{subfigure}
	\begin{subfigure}[b]{\textwidth}
		\includegraphics[width=\textwidth]{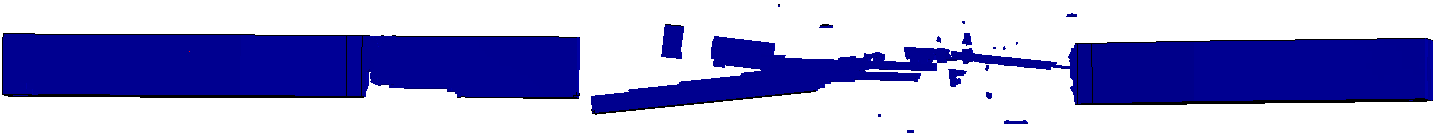}
		\caption{Simulation $0 \, ^\circ$ UD}
		\label{fig:BB_0_sim}
	\end{subfigure}
	\caption{Comparison of fracture patterns}
	\label{fig:BB_all}
\end{figure}

\paragraph{Parameter identification}	
The introduced damage model is pa\-ra\-met\-rised with tensile tests for the two lay-ups presented in \autoref{tab:MatConfig}, the complete material card used for the simulations can be found in \autoref{tab:MatCard}. The simulations are fitted to the effective experimental stress-strain curves.
The strains were both measured with strain gauges and calculated from crosshead displacement. The results of the strain gauges were used to identify the initial stiffness and compared to the simulated stiffness. Because of the influence of delaminations and local damage on the strain gauges, these results can not be evaluated for higher strains. For this reason the whole stress-strain curves, including failure, which are plotted in \autoref{fig:Paramet}, are evaluated with the strains based on the crosshead displacement.\\
The  $0 \, ^\circ$ tensile test is an exception, there slippage of the test specimens in the clamping area could not be excluded, so this test is completely evaluated with strain gauges.\\
For parameter identification, the specimens were modelled with fully integrated solid elements, for each ply three elements were used in thickness direction, delamination is modelled with cohesive zone elements with zero thickness.\\
Cohesive zone elements account for mode I (tensile failure) and mode II failure (shear failure). For each mode, a strength and a fracture toughness are needed, here the same values as for the associated solid elements were used. For mode II the longitudinal shear strength showed to be more useful than the traversal one.\\
In order to reach comparability of the simulated and experimental results, the whole specimen was modelled and stresses were calculated as the quotient of force and cross-sectional area in the simulation. Strains measured with strain gauges are compared to elemental strains in the same location and strains calculated from cross-head displacement are compared the quotient of boundary node displacement and initial free length of the specimen.
\begin{table}[H]
	\centering
	\begin{tabular}{l r r r r r r}
		\toprule
		Para-&$E_\parallel$&$E_\bot$&$\nu_{\bot \parallel}$&$\nu_{\bot \bot}$ & $G_{\bot \parallel}$ & $G_{\bot \bot}$\\
		meter &$\mathrm{N \, m}^{-2}$&$\mathrm{N \, m}^{-2}$&$-$&$-$ &$\mathrm{N \, m}^{-2}$ &$\mathrm{N \, m}^{-2}$\\
		Value  & $1.21\;10^{11}$ & $9.24 \;10^9$ & $0.015$ & $0.37$  & $1.35\;10^{10}$ & $2.92\;10^{9}$\\ 
		\midrule
		
		Para-&$\Gamma_{\parallel, C}$ & $\Gamma_{\parallel, T}$  & $\Gamma_{\bot, T}$ &$\Gamma_{\bot \bot}$&$\Gamma_{\bot \parallel}$&$\rho$\\
		meter &$\mathrm{N \, m}^{-1}$&$\mathrm{N \, m}^{-1}$&$\mathrm{N \, m}^{-1}$&$\mathrm{N \, m}^{-1}$&$\mathrm{N \, m}^{-1}$&$\mathrm{kg \,m}^{-3}$\\
		Value  & $4.0\;10^{7}$& $4.5\;10^{7}$& $2.5\;10^{3}$& $4.0\;10^{3}$& $4.0\;10^{3}$ & $1530$\\ 
		
		\midrule
		
		Para-&$X_{\parallel,C}$&$X_{\parallel,T}$&$X_{\bot,C}$&$X_{\bot,T}$&$X_{\bot \parallel}$&$\varepsilon_\mathrm{eff}$\\
		meter &$\mathrm{N \, m}^{-2}$&$\mathrm{N \, m}^{-2}$&$\mathrm{N \, m}^{-2}$&$\mathrm{N \, m}^{-2}$ &$\mathrm{N \, m}^{-2}$ & $-$\\
		Value  & $6.5\;10^{9}$ & $1.79 \;10^9$ & $1.3 \;10^8$ & $4.2 \;10^7$  & $7.4\;10^{7}$ & 0.16\\ 
		
		\bottomrule
	\end{tabular}
	\caption{Material Card for Pinho damage model}
	\label{tab:MatCard}
\end{table}
The stiffnesses, strengths and Poisson's ratios were identified from the uni-directional tensile tests with fibre angles of $0 \, ^\circ$ (\autoref{fig:UD0}), $90 \, ^\circ$ (\autoref{fig:UD90}) and $45\, ^\circ$ (\autoref{fig:UD45}). 
In order to address the scattering observed in experiments,
the simulation was fitted to the mean values of the stiffnesses and strengths.\\
For the  $\pm 45 \, ^\circ$ multilayer tensile test (\autoref{fig:ML+-45}) the characteristic nonlinear shear stress-strain behaviour, which was described by Paepegem et al.\cite{Paepegem2006modelling},
was observed in experiments. The experimental stress-strain curve is used as a master curve for the shear plasticity in the material model.
\\
The whole damage model is completely para\-metrised with the presented tensile tests and therefore the result of the simulation of the ML $0\,^\circ/90 \, ^\circ$ tensile test  gives a first validation. For this simulation no additional information was added to the material model, but the simulation results in \autoref{fig:ML0-90} show a good prediction for this test.\\ 
One adaptation that was made according to the material layup are the criteria for element deletion. In $45 \,^ \circ$ and $90 \,^ \circ$ UD tensile tests (\autoref{fig:UD45}) elements are deleted after reaching a damage value of $d=1$ for an arbitrary mode. For all the other simulations element deletion was disabled for the matrix mode and elements were deleted either after reaching the fibre criterion $d_\parallel = 1$ or the strain criterion $\varepsilon_\mathrm{eff}=0.16$. 
\\
In addition to the stress-strain curves, the fracture behaviour can be evaluated with the comparison of the fracture patterns \autoref{fig:BB_all}.
In the  UD $90 \,^ \circ$ tensile tests in experiment (\autoref{fig:BB_90_exp}) and simulation (\autoref{fig:BB_90_sim})  just a straight single crack parallel to the fibres occurred. In contrast to that, in the  multilayer tests (\autoref{fig:BB_0-90_exp} and \autoref{fig:BB_0-90_sim}) more extensive damage, e.g. fibre pull-out and delaminations occurred. The most extensive damage was observed in the UD $0 \,^ \circ$ tensile tests. In the simulation first a single crack grew perpendicular to the fibres. Due to the high energy release of the initial crack, multiple smaller cracks grew parallel to the fibres, which lead to the characteristic patterns in \autoref{fig:BB_0_exp} and \autoref{fig:BB_0_sim} with multiple slender fragments.\\
The comparison of the fracture patterns in experiment and simulation shows a good agreement.
This is a good indicator, that the identified fracture toughnesses fit the energy release during fracture well.

	\subsection{Modelling of the heat affected zone}
	\paragraph{HAZ model with damage parameters}
A possible modelling approach for initial damage introduced by the laser cutting process is the utilisation of damage parameters of the introduced material model. I.e., in contrast to initially intact elements pre-damaged elements have non-zero damage parameters $d_0$ for different modes. For this modelling approach, mode dependent damage parameters have to be identified as well as the action plane presented in \autoref{fig:ActPlane} for matrix failure.\\
Inter-fibre damage, as the evaporated or pyrolysed matrix, can be modelled with the matrix damage variable $d_\bot$, the fibre damage variable $d_\parallel$ is only needed if fibre fractures occur. For this case it has to be noted, that stresses in matrix direction are degraded by both damage variables, as presented in \autoref{tab:DamageVariables}, so an interaction must be considered.
Concerning the direction of the action plane, it showed to be a useful assumption to set the action plane perpendicular to the specimen plane. In this way, the degraded normal direction is in the specimen plane.\\
A restricting side effect of this modelling approach using damage parameters is the fixed relation of stiffness, strength and toughness reduction. This leads to a softening of the damaged material, i.e. the fracture strain increases with increasing damage, which can delay the failure of a structure.
\\
Another effect was observed in simulations with this modelling approach as well: regardless of the parametrisation,  pre-damaged specimens always overestimated the effective stiffness reduction and so the resulting strength. In fact, the experimentally observed increase of the maximum effective stress was exceeded by at least factor two in the simulation, as presented in \autoref{fig:Error}.\\
An explanation for this deviation can be found in the definition of damage parameters, that are used to model the macroscopic effects of micro cracks. But in the HAZ there occur also additional damage phenomena like chemical decomposition of the epoxy, which apparently can not be modelled exclusively with damage variables.\\
Since the damage parameters alone are not suitable to model the thermally degraded material, this approach was no further pursued, but an alternative phenomenological approach is introduced.

\begin{figure}[!h]
	\def\svgwidth{\columnwidth}	
	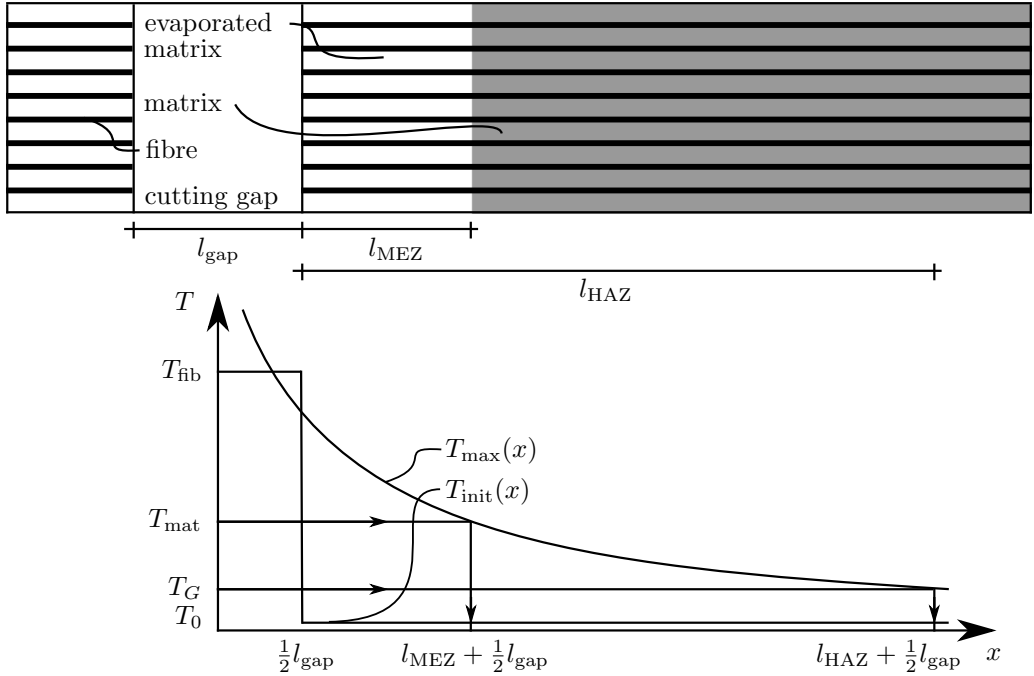
	\caption{Modelling approach for HAZ}
	\label{fig:skizze}
\end{figure}

\paragraph{HAZ model with material parameter reduction}
An alternative approach is the direct reduction of the material parameters. Compared to the first approach, this allows for a free ratio between the reduction factors for stiffness, strength and fracture toughness.\\
A similar modelling approach was introduced in \cite{stock2016}, where the maximum temperature in a material point is identified as the key parameter for the reduction. A dependence on the interaction time was neglected.  Under this premise, tensile specimens were pre-heated to different temperatures from $20 \,^\circ\mathrm{C}$ to $500 \,^\circ\mathrm{C}$ and subsequently tested.
with increasing temperature, the stiffness and tensile strength both decreased due to the matrix degradation and evaporation. The measured temperature dependent material parameter reduction of the tensile specimens were transferred to the material points in the HAZ with equal maximum temperature during the cutting process.\\
Following this approach, the HAZ model requires the identification of reduction factors of the material parameters and their spatial distribution.
\paragraph{Size of the HAZ}
\label{sec:lHAZ}
No matter which modelling approach is used, the dimensions of the HAZ are crucial input variable for the model. Two indicators for the width can be measured in micro sections of the cutting edge: first the width of the gap $l_\textrm{gap}$ and second the nominal length $l_\textrm{MEZ}$ of the matrix evaporation zone. According to \cite{freitag2017energietransportmechanismen}, $l_\textrm{MEZ}$ can be measured layer-wise as the mean width dividing the evaporation area $A_\mathrm{MEZ}$ with the layer thickness $h_\mathrm{layer}$ for the left and the right side individually, see \autoref{fig:Schliff}.\\
The HAZ does not only contain those areas where the matrix is evaporated, but is defined as the whole area where the mechanical properties of the material are changed, as presented in \autoref{fig:skizze}. Those changes, caused by chemical decomposition and micro structural transformations, start with the glass transition temperature of the epoxy. \\
An upper limit for this length can be calculated from one-dimensional heat conduction, as presented by Weber et al. \cite{weber2011minimum}. 
For the infinite one-dimensional heat conduction, a coordinate system as presented in \autoref{fig:Schliff} is used, whereby $x$ is the direction of heat transport and the origin is at half height in the middle of the gap.
For layers perpendicular to the cutting edge this is a good approximation because of the much higher conductivity in fibre direction than perpendicular to fibre (factor 19.9 with Chamis' rule of mixture \cite{chamis1983simplified}, parameters in \autoref{tab:MatPara}). The 1D heat conduction is calculated for the perpendicular middle-layers, since the medium gap width at half height $l_\mathrm{gap}(y=0)$ is a good representation for the whole composite.\\
\begin{table}[h]
	\centering
	\begin{tabular}{p{2.7 cm}rrrr} 
		\toprule
		Parameter& Unit & Matrix & \multicolumn{2}{c}{Fibre} \\ 
		&  &  & (longitudinal) &  (radial)\\ 
		\midrule 
		Heat \mbox{conductivity} $\lambda $&$ \mathrm{W(m\,K)}^{-1}$& 0.21 & 14 &0.75 \\
		Evaporation temperature &$ \mathrm{K}$ & $T_\mathrm{mat}=800$ \cite{klotzbach2011laser} & \multicolumn{2}{c}{\qquad $T_\mathrm{fib}=3900$ \cite{weber2011minimum} } \\
		Glass transition temperature $T_G$&$ \mathrm{K}$ & $413.15$ \cite{SGL2020} & \multicolumn{2}{c}{--} \\
		Reference temperature $T_0$&$ \mathrm{K}$ &  \multicolumn{3}{c}{$298.15$} \\
		\bottomrule
	\end{tabular}
	\caption{Thermal Material Parameters}
	\label{tab:MatPara}
\end{table}
The assumption for the initial state is, that in the whole gap width the fibre sublimation temperature $T(\mid x \mid < 0.5 \, l_\mathrm{gap}, t=0)=T_\textrm{fib}$ is reached. In the remaining area, the temperature equals the reference temperature $T(\mid x \mid \ge 0.5 \, l_\mathrm{gap}, t=0)=T_0$, as presented with the step function in \autoref{fig:skizze}. \\
As stated in \cite{mueller2001Grund}, for each time $t$ the temperature profile can be calculated as
	\begin{equation}
		T(x,t)=
		\frac{T_\mathrm{fib}-T_0}{2\sqrt{\pi \lambda t}} e^\frac{-x^2}{4 \lambda t} + 
		T_0
		\label{eq:Txt}
	\end{equation}
	using the heat conductivity $\lambda$. For the determination of the HAZ, not the whole temperature profile, but only the maximum temperature for a material point is of interest. This is calculated with the partial derivative
	\begin{equation}
		\frac{\partial T}{\partial t} =0
	\end{equation}
	as
	\begin{equation}
	T_\textrm{max}(x)=T_\textrm{0}\Big(1+ \frac{1}{\sqrt{2 \pi e}} \frac{T_0-T_\textrm{fib}}{T_\textrm{fib}}  \frac{l_\textrm{gap}}{x}\Big).
	\label{eq:Tmax}
	\end{equation}
Here, it has to be considered, that the initial temperature profile is approximated with \autoref{eq:Txt} to conserve the total energy. Consequently, there are deviations in the temperature profile $T_\mathrm{max}$ for small values $x$ (see \autoref{fig:skizze}, $T_\mathrm{max}(x=0.5 l_\mathrm{gap})\ne T_\mathrm{fib}$). Nonetheless the equation is evaluated in a distance where the errors are much smaller.\\
Using the inverse of {} \autoref{eq:Tmax}, $l_\mathrm{MEZ}$ and $l_\mathrm{HAZ}$ can be calculated as the maximum length where $T_\mathrm{mat}$ and $T_G$ were exceeded.\\
	\paragraph{Identification of HAZ parameters}
	For the mechanical HAZ model, reduction parameters have to be identified.\\
	In \cite{stock2016} thermally degraded CFRP specimens were tested. 
	The specimens are stored in an oven several minutes in order to reach a homogeneous temperature field and thus a homogeneous damage state in the whole specimen, which is the fundamental precondition for the assignment of macroscopic parameter reduction in the specimen to the local material point.
	The reduction factors of Young's modulus and tensile strength at $500 \,^\circ \mathrm{C}$, so above the matrix decomposition temperature, were identified as $f_\textrm{Young}=22\,\%$ and $f_\textrm{strength}=68 \, \%$. 
	These values 
	were assigned as reduction factors at the cutting gap and are used in this contribution too, as both materials are CFRP with an epoxy matrix and Stock et al. could correctly predict the stress distribution of laser cut open hole specimens with this approach.\\	
	The fracture toughness reduction factor is calculated depending on the other factors $f_\textrm{tough}=1-(1-f_\textrm{Young})(1-f_\textrm{strength})$,
	so 
	a decreased stiffness as well as a decreased strength leads to a decrease of the element deletion strain $\varepsilon^2$.
	Although this is a reasonable assumption, still further research of the relation between the reduction factors of those three parameters may be interesting.
	\\
	Identical reduction factors are applied for fibre and matrix mode, since the factors were measured as effective values for the composite. \\	
	For the spatial distribution of the reduction factors, different variants were tested in a sensitivity analysis. As a result, the influence of the shape of the spatial distribution on the fracture behaviour was much smaller than that of the reduction factors $f_\textrm{strength}$ and $f_\textrm{Young}$ and the maximum length $l_\mathrm{HAZ}$. For this reason a linear distribution as the simplest approach is used.\\
	An alternative is to utilisation the shape of the temperature field, but this does not seem reasonable to the authors for two reasons. First, the 1D temperature field is only correct if the fibres are perpendicular to the cutting edge, so only for half of the layers under certain angles. This allows for a good prediction of the upper limit of $l_\mathrm{HAZ}$, but is not a precise depiction of the 3D temperature field. Second, the modelling strategy takes a conservative approach, but the convex temperature function might underestimate the damage intensity compared to a linear function. \\	
	Since the cohesive zone elements for initially intact material were parametrised with the mechanical parameters of their associated elements, this approach is used for initial delaminations in HAZ as well, hence, the same degradation factors are applied to the cohesive zone elements as to the associated solid elements.
	\paragraph{FE Modelling}
	As for previous tensile test models, each layer is modelled with three elements in thickness direction and between the layers there are zero-thickness cohesive zone elements. The load is applied as a prescribed displacement to all surface nodes in one of the two clamping areas, the other degrees of freedom are locked for these nodes. In the area of the hole, there are the smallest elements with 100 elements in its circumference and a minimum element size of $0.64 \, \mathrm{mm}$ in radial direction. This fine mesh is necessary to guarantee that there are multiple elements over the length of the HAZ.\\
	The reduction factors introduced above are determined for all nodes, the element reduction factor is defined as the maximum of the eight assigned nodes.

	\section{Results and discussion}
\begin{figure}[!h]
	\centering
	\def\svgwidth{8 cm}
	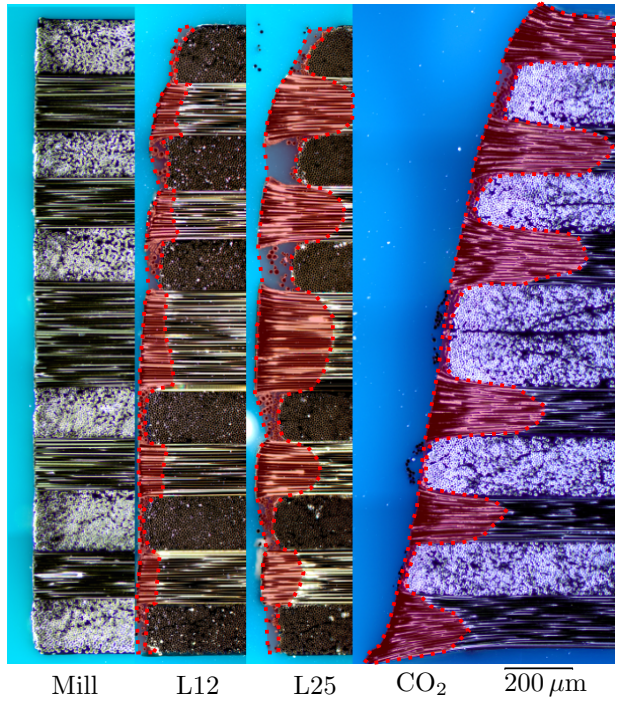
	\caption{Comparison of the evaporated matrix for the different cutting technologies.}
	\label{fig:VglCutConf}
\end{figure}

\paragraph{1D thermal conduction}
For the three laser cutting parameter configurations introduced in \autoref{tab:CutConf1}, micro sections were prepared, which are presented in \autoref{fig:VglCutConf}. In this images $l_\mathrm{gap}$ and $l_\mathrm{MEZ}$ were measured, the values are given in \autoref{tab:CutConf2}.

\begin{table}[!h]
	\centering
	\begin{tabular}{p{3.5cm} rrrr} 
		\toprule
		Parameter & Unit & L12 & L25 & CO$_2$ \\
		\midrule
		Gap width                      & $\mu \mathrm{m}$      & 50 & 170 &  450* \\
		Matrix evaporation (measured)  & $\mu \mathrm{m}$      & 67 & 200 &  351 \\
		Matrix evaporation (calculated)& $\mu \mathrm{m}$      & 61 & 208 &  351 \\
		HAZ width                      & $\mu \mathrm{m}$      & 351& 1194 & 2016\\
		\bottomrule
	\end{tabular}
	\caption{HAZ characterisation for the examined Laser cutting configurations.\\ 450*
		here $287\,\mathrm{\mu m}$ were used in calculation to fit the width of evaporated matrix. For
		small gaps it is assumed that the heat is mainly conducted into the material, this
		assumption is no longer correct for broad gaps, where the hot particle gas can freely
		leave to the top.
	}
	\label{tab:CutConf2}
\end{table}
As the comparison of measured and calculated $l_\mathrm{MEZ}$ in \autoref{tab:CutConf2} shows, the calculation of $l_\mathrm{MEZ}$ based on 1D thermal conduction works precisely for narrow, fibre laser cut gaps. The deviation between measured and calculated matrix evaporation zone length was less than $0.1 \, \mathrm{mm}$ for both fibre laser configurations.\\ 
For broader gaps the 1D approximation is less accurate. Through the broad cutting gap hot particle gas can move freely and thus transport the heat away, which results in  an overestimation of $l_\mathrm{MEZ}$ in the calculation.
Hence, for $\mathrm{CO}_2$ laser cutting the gap width was changed in order to fit the matrix evaporation length.

\begin{figure}[!h]
	\centering
	\def\svgwidth{0.6 \columnwidth}	
	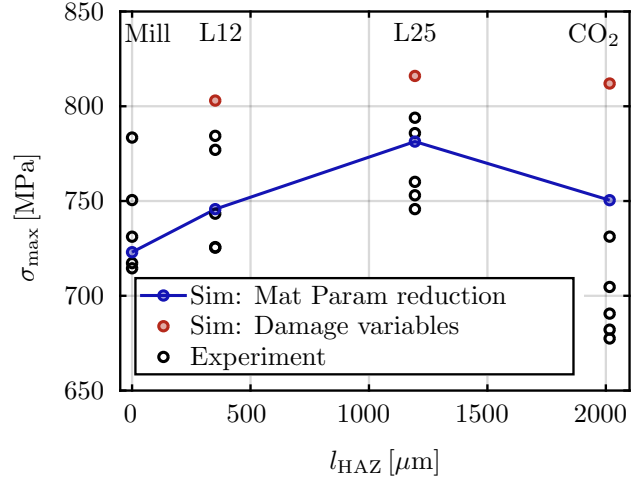
	\caption{Relation between HAZ width $l_\mathrm{HAZ}$ and maximum effective stress $\sigma_\mathrm{max}$}
	\label{fig:Error}
\end{figure}

\begin{figure}[H]
	\def\svgwidth{\columnwidth}	
	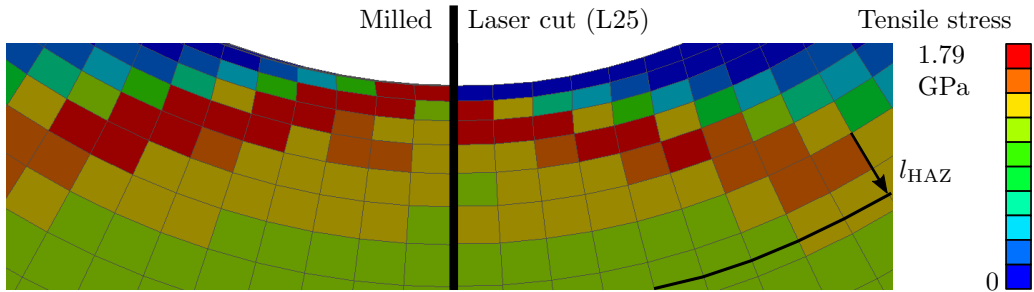
	\caption{Comparison of stress fields for milled (left) and laser cut (right) specimens in simulation.}
	\label{fig:Contour}
\end{figure}

\begin{figure}[H]
	\def\svgwidth{\columnwidth}	
	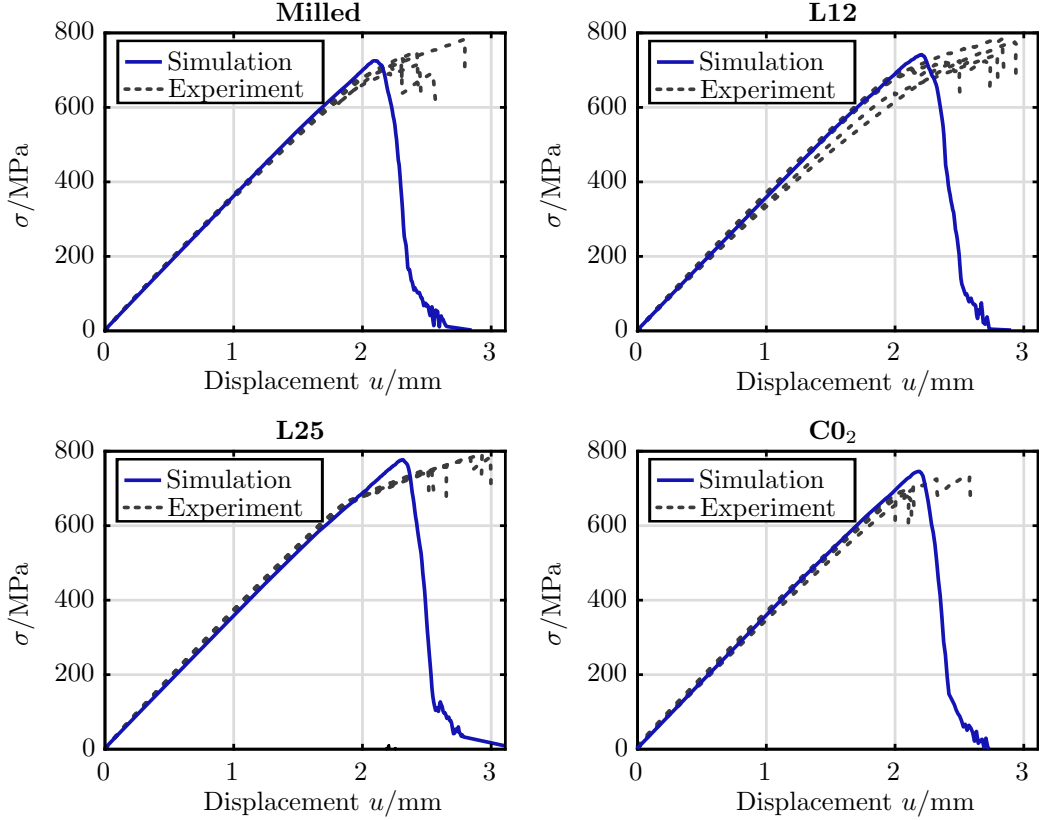
	\caption{Open hole test results for different cutting technologies}
	\label{fig:KW_1loch}
\end{figure}

\begin{table}[!h]
\centering
\begin{tabular}{p{2 cm} rr rrr} 
	\toprule
	Cutting& \multicolumn{2}{c}{Experiment} & \multicolumn{3}{c}{Simulation} \\
	Technology& \multicolumn{2}{c}{$\sigma_\mathrm{Max}$}& \multicolumn{3}{c}{$\sigma_\mathrm{Max}$}\\
	& Mean & $s$&  Abs  & Err & Err\\
	Unit& $\mathrm{MPa}$&$\mathrm{MPa}$&$\mathrm{MPa}$&$\mathrm{MPa}$ & \%\\
	\midrule
	Milled & $739.6$ & $3.84$ &  $723.0$ & $ -16.6$ & $ -2.24$\\ 
	L12    & $751.0$ & $3.72$ &  $745.7$ & $ -5.30$ & $ -0.70$\\ 
	L25    & $767.8$ & $2.74$ &  $781.4$ & $ +13.6$ & $ +1.77$\\ 
	CO$_2$ & $689.2$ & $1.51$ &  $750.4$ & $ +53.2$ & $ +7.63$\\ 
	\bottomrule
\end{tabular}
\caption{Tensile Test Results, five specimens for each cutting method, Stresses defined as Quotient of force and smallest diameter at the hole in order to balance deviations in plate thickness and hole diameter}
\label{tab:Results}
\end{table}
\paragraph{Open hole tensile test}
As a result of the open hole tensile test, the maximum stress $\sigma_\mathrm{max}$ is evaluated. The stresses in this test are nominal stresses,  i.e. the quotient of force and the smallest cross-sectional area at the hole. The experimental and simulated values of $\sigma_\mathrm{max}$ are given in \autoref{tab:Results}. In \autoref{fig:Error} these values are plotted over $l_\mathrm{HAZ}$ in order to visualise the influence of the HAZ on $\sigma_\mathrm{max}$	\\
	The experimental results show, that an increased $l_\mathrm{HAZ}$ leads to 
	increased strength $\sigma_\mathrm{max}$ for remote laser cut specimens (L12 and L25) compared to milled specimens, which is caused by the stiffness reduction in the HAZ, that reduces the notch effect, as stated in \cite{zaeh2017peak}.\\
CO$_2$ laser cut specimens have the largest $l_\mathrm{HAZ}$, but this does not lead to a further increased $\sigma_\mathrm{max}$. The CO$_2$ laser cut specimens have the smallest tensile strength. An explanation for this can be found in the reduced cross section, which dominates over the smaller stress concentration.
Another factor is the smaller spot velocity of the CO$_2$ Laser, which increases the interaction time and so possibly the damage intensity.
\\
In the comparison of numerical and experimental results for milled and fibre laser cut specimens, the simulated $\sigma_\mathrm{max}$ are in the range of variation of the experiments. Only for $\mathrm{CO}_2$ laser cut specimens the strength $\sigma_\mathrm{max}$ is overestimated.
This shows that the presented HAZ material model is able to predict the general trend observed in the experiments, particularly for fibre laser cut specimens. For CO$_2$ laser cut specimen the tensile strength is overestimated in the simulation, 
but the trend of a decreased strength $\sigma_\mathrm{max}$ could be reproduced.\\
The whole stress-displacement curves are plotted in \autoref{fig:KW_1loch}. In experiments the displacements $u$ are measured with an extensometer, in the simulation the node displacements are used.\\
The stiffness reduction at the cutting gap, which has a significant impact on the increased $\sigma_\mathrm{max}$ of the fibre laser cut specimens is also visible in the comparison of the stress fields of milled and fibre laser cut specimens in \autoref{fig:Contour}. The stiffness of the elements at the cutting gap of the laser cut specimen is reduced. Therefore, the maximum stress does not occur at the edge, as in the milled specimen, but at some distance to the edge, which delays failure initiation.

	\paragraph{Discussion}
	Even though the introduced HAZ modelling approach is able to reproduce the trend of  the $l_\mathrm{HAZ}$-$\sigma_\mathrm{max}$-relation, the limitations of the model and reasons for the deviation need to be discussed.
	\\
	The largest deviations occurred for the CO$_2$ laser cut specimens which have the largest HAZ. Another difference between fibre and CO$_2$ laser cutting is the spot velocity, which is much higher for cutting with fibre laser. Consequently the interaction time for CO$_2$ is higher, which causes the larger HAZ. As stated by Chippendale \cite{chippendale2014numerical}, the matrix degradation is time-dependent, so with a higher interaction time also the intensity of the thermal damage might increase, even though the rate dependency was only observed for much higher interaction times than those of the cutting process.\\
	Possible time-dependent effects are not yet considered, as the introduced modelling approach only calculates the maximum temperature and not the interaction time of this temperature.
	Furthermore, the 1D thermal calculation uses simplifications: effects like material evaporation in the cutting gap and interaction with the hot process gas, the heat input by the laser and the influence of rest times between the cutting cycles can not be accounted in this calculation.
	In addition, this approach only gives an upper boundary for $l_\mathrm{HAZ}$ and can not be used for a layer-wise evaluation.
	
	\section{Conclusions and future work}
	In this article, investigations on the influence of laser cutting on the failure of CFRP were presented. For this purpose an experimental and numerical analysis of open hole specimens produced with different cutting technologies were performed. Based on a parametrised damage model, a HAZ modelling approach was introduced and demonstrated.\\
	The complex interactions between different degradation processes and their mechanical effects could be reproduced. Compared to thermally undamaged milled specimens, fibre laser cut specimens showed a higher tensile strength. This shows that the aim of a cutting process optimisation does not have to be a minimal HAZ, as the highest tensile strength was observed for a medium HAZ.\\	
	For CO$_2$ laser cut specimens with a larger HAZ a reversed trend was observed, here the tensile strength is decreased. The simulation was able to predict the trend of the tensile strength to first increase for larger HAZ and later decrease again.\\ 
	For future investigations, the HAZ modelling approach shall be refined. The temperature during the cutting process can be calculated in a cutting process simulation which allows for a more accurate determination of the HAZ width for the different layers. Regarding the time-dependency of the matrix degradation process in the thermal simulation and calculating the material parameter reduction based on the degradation, the mechanical simulation can be improved as well.\\
	The influence of the HAZ zone is not only limited to the static fracture behaviour, but there are also additional effects in the fatigue behaviour, which the authors intend to determine in their future research.
	
	\section*{Acknowledgement}
	This work was supported by the Deutsche Forschungsgemeinschaft (DFG, KA 3309/6-1 and ZI 1006/12-1). The authors gratefully acknowledge the support. 
	
	\section*{References}
	
	\bibliography{Bib}
	\bibliographystyle{elsarticle-num}
		
\end{document}